# Multimode waveguide crossing based on square Maxwell's fisheye lens

S. HADI BADRI[1], H. RASOOLI SAGHAI[2], HADI SOOFI[3]

[1]Department of Electrical Engineering, Azarshahr Branch, Islamic Azad University, Azarshahr, Iran
[2]Department of Electrical Engineering, Tabriz Branch, Islamic Azad University, Tabriz, Iran
[3]School of Engineering- Emerging Technologies, University of Tabriz, Tabriz 5166616471, Iran
 sh.badri@iaut.ac.ir

**Mode-division multiplexing (MDM) is an emerging large-capacity data communication technology utilizing orthogonal guiding modes as independent data streams. One of the challenges of multimode waveguide routing in MDM systems is decreasing the mode leakage of waveguide crossings. In this article, a square Maxwell's fish-eye lens as waveguide crossing medium based on quasi-conformal transformation optics is designed and implemented on the silicon-on-insulator platform. Two approaches were taken to realize the designed lens: graded photonic crystal and varying the thickness of Si slab waveguide. Three-dimensional numerical simulations show that the designed multimode waveguide crossing has an ultrawide bandwidth from 1260 to 1675 nm with a compact footprint of only 3.77×3.77 µm$^2$. For the first three transverse electric modes (TE0, TE1, and TE2), the designed waveguide crossing exhibits an average insertion loss of 0.24, 0.55, and 0.45 dB and the crosstalk of less than -72, -61, and -27 dB, and a maximum return loss of 54, 53, and 30 dB, respectively. The designed waveguide crossing supports low distortion pulse transmission with a high fidelity factor of 0.9857. Furthermore, proposed method can be expanded to design waveguide crossings with even higher number of supporting modes by increasing the size of the lens.**

## 1. Introduction

Multiple precise wavelength sources required in costly wavelength-division multiplexing (WDM) systems and their limited bandwidth scalability have limited their use to large-scale systems. However, the cost-effective MDM technology can carry multiple data channels corresponding to separate guiding modes with a single wavelength source [1]. Compared to the single mode communication systems, MDM systems can considerably increase the data communication capacity. Waveguide crossings with low inter-mode crosstalk and small footprint are essential in MDM systems. Multimode waveguide crossings have been studied with different methods [2]. Multimode interference (MMI) crossings are based on creating self-imaging for the guiding modes. Optimizing the width and length of the MMI crossing to support more than two modes is very difficult and, therefore, only two-mode waveguide crossings based on MMI have been reported [3, 4]. The insertion loss and crosstalk lower than 0.6dB and -24dB, respectively, in a 60nm bandwidth and 4.8×4.8 µm$^2$ footprint have been achieved for two-modes for MMI crossing [4]. It has been suggested to convert higher-order modes into the fundamental mode and use regular single-mode intersections in order to increase the number of supporting modes. In this method, to support N modes, N×N single mode intersections are required. Therefore, the crossing of two waveguides based on this method, supporting two [5] and three [6] modes, has a large footprint of 21×21 and 34×34µm$^2$, respectively. Recently, researchers have proposed novel applications for classic gradient index (GRIN) lenses such as Luneburg [7, 8], MFE [9, 10], and Eaton [11] lenses. Multimode waveguide crossings based on Maxwell's fisheye (MFE) lens have been proposed. The MFE lens in these crossings has been implemented by different methods [12-14].

In this article, a compact 2×2 multimode waveguide crossing is theoretically designed and numerically evaluated. The presented device exhibits a bandwidth of 415 nm (from 1260 to 1675 nm) with crosstalk levels lower than -27 dB. Moreover, the device has a very small footprint of less than 4×4 µm$^2$ making it ideal for the emerging nanophotonic integrated circuits. The design is based on MFE lens, and its geometrical parameters are determined by the transformation optics (TO). The refractive index profile of the circular MFE lens is:

$$n_{lens}(r) = \frac{2 \times n_{edge}}{1+(r/R_{lens})^2} \quad , \quad (0 \leq r \leq R_{lens}) \quad (1)$$

where $n_{edge}$ is the refractive index of the lens at its edge, $r$ is the radial distance from the center of the lens and $R_{lens}$ is the radius of the lens. To the best of our knowledge, this is the first study that investigates the performance of the square MFE lens as the waveguide crossing. We have presented photonic crystal waveguide crossings based on the circular MFE lens [9, 13] previously and we have recently transformed the circular MFE lens into hexagonal and octagonal lenses to design 3×3 and 4×4 waveguide crossings [14]. The concept of square MFE lens arises due to an intrinsic limitation of conventional circular MFE lens for silicon-on-insulator waveguides. In order to design a waveguide crossing for SOI waveguides (with an effective refractive index of approximately 2.1) by the circular MFE lens, the $n_{edge}$ should be 2.1 to reduce reflection at the interface of the lens and the waveguides. Based on Eq. (1), the refractive index of the circular lens at its center would be 4.2 which cannot be implemented by conventional materials and utilizing metamaterials reduces the bandwidth of the crossing severely. In order to increase the refractive index of the lens's edge without doubling the refractive index at its center, we employ TO for transforming the circular lens into the square one. By this modification, the reflection can be minimized while keeping the broad operation spectrum intact. Moreover, the proposed transformed lens offers an additional advantage of matching the flat wavefront of waveguide modes with the wavefronts of the square MFE lens at its boundary which was not possible for circular MFE lens. This feature minimizes the reflection and enhances the characteristics of the lens even more.

## 2. Designing square Maxwell's fisheye lens

Transforming the given geometry, virtual domain, into a new geometry, physical domain, may require anisotropic permittivity or permeability with very high or low values [15]. In quasi-conformal transformation optics (QCTO), by imposing some limitations, complexities of the materials required for the implementation of the physical domain can be minimized [16]. The two quadrilateral domains, with the same conformal module, can be mapped into an intermediate domain which is a rectangular with the same conformal module. Conformal module is the ratio of the lengths of the two adjacent sides of a domain [17, 18].

In this work, the circular MFE lens in the virtual domain is transformed to a square lens in the physical domain [8]. The first step in QCTO is to generate an orthogonal grid, i.e., grid lines are orthogonal to each other, in the virtual and physical domains [18]. The orthogonal grid is generated by solving the Laplace equation. Boundary orthogonality is achieved by applying Dirichlet-Neumann boundary conditions. Knowing that inverse of a conformal mapping is conformal, two domains with the same conformal module, M, can be mapped onto each other conformally by mapping them onto an intermediate domain. The intermediate domain is a rectangle with the same conformal module, M [17]. Conformal module is the ratio of the lengths of the two adjacent sides of a domain. As shown in Fig. 1, in order to ease the constraints on orthogonal grid generation, four corner-like protrusions are added to the lens in the virtual domain [19]. The material properties of the physical domain are described with [20]:

$$\varepsilon = \frac{J\varepsilon'J^T}{|J|} \quad , \quad \mu = \frac{J\mu'J^T}{|J|} \quad (2)$$

where $\varepsilon$ and $\mu$ are transformed permittivity and permeability in the physical domain, respectively. $J$ is the Jacobian transformation matrix that relates the coordinates between the virtual and physical domains, and the spatial distribution of permittivity, $\varepsilon'$, and permeability, $\mu'$, in the virtual domain is given by

$$\mu' = 1 \quad , \quad \varepsilon' = n_{vir}^2(r) = n_{lens}^2(r) \quad (3)$$

In our design, $n_{edge}$ is chosen as 1.25 and the radius of the lens is $R_{lens}$=5.58 μm.

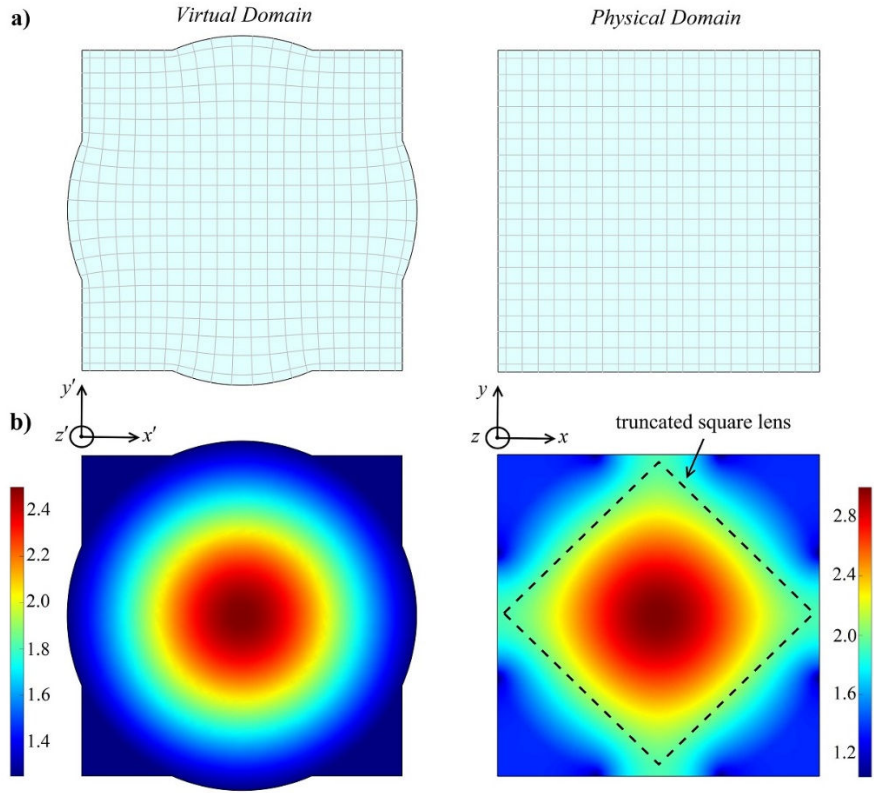

Fig. 1. a) QCTO mapping of virtual to physical domain. b) refractive index profile of virtual and physical domains. The truncated square MFE lens is shown with dashed lines.

The calculated refractive index of the square lens is shown in Fig. 1(b). We have truncated the square lens in order to match the refractive index of the square lens's sides to the refractive index of the input/output waveguide cores [12]. The truncated square lens, which is used as the crossing medium, is shown with dashed lines in Fig. 1(b). The length of its sides is 3.77μm. Therefore, the footprint of the square MFE lens is reduced by 54% compared to the circular MFE lens.

## 3. Implementation of the square MFE lens

We have implemented the designed lens on SOI platform with two methods: GPC and varying the thickness of Si slab waveguide. The performance of the two-dimensional (2D) lens implemented with GPC was investigated by finite element method (FEM). Since finite-difference time-domain (FDTD) method consumes less memory, the performance of the three-dimensional (3D) lens implemented with varying the thickness of Si slab waveguide was investigated by this method.

### 3.1 Graded photonic crystal

Based on effective medium theory (EMT), GPC can be used as low-loss and broadband metamaterial. GPC is a gradually modulated photonic crystal with spatially varying geometrical or material properties [21]. The electromagnetic properties of the metamaterials originate from its sub-wavelength structures, not the constituent materials [22]. In our design, the radius and position of the subwavelength rod-shaped inclusions were controlled to implement the designed refractive index profile. The approximation method for mapping a 3D model to a 2D model were based on simple area averaging of host and inclusion materials by ignoring the evanescent mode in claddings [12]. The approach in implementing the GRIN medium with GPC is to divide the GRIN medium into cells of appropriate size and shape. Then each cell's average refractive index, $n_{eff,ij}$, is approximated by placing a rod-shaped inclusion with a radius of $r_{rod,ij}$ at the center of the cell, i.e., $x_i$ and $y_j$. The square MFE lens is divided into square cells.

The TE mode, where the electric field is parallel to the rods axis, is excited in 2D simulations. For this mode, the effective permittivity of the GPC with cylindrical rods in a host material is approximated by volume averaging theory [23]:

$$\varepsilon_{eff} = f_{rod}\varepsilon_{rod} + (1-f_{rod})\varepsilon_{host} \qquad (4)$$

where $\varepsilon_{rod} = n_{rod}^2$, $\varepsilon_{host} = n_{host}^2$, and $\varepsilon_{eff} = n_{eff}^2$ are the permittivity of the rods, host, and effective medium of the cell, respectively. And the filling ratio of the rods is $f_{rod,ij} = \pi r_{rod,ij}^2 / A_{ij}$ where $A_{ij}$ is the area of the ij-th cell. The radius of the rod in square unit cells is given by

$$r_{rod,ij} = a_{GPC}\sqrt{\frac{(\varepsilon_{eff,ij}-\varepsilon_{host})}{\pi(\varepsilon_{rod}-\varepsilon_{host})}}$$

where $a_{GPC}$ is the lattice constant of the square unit cell. Two implementations of the GPC-based lens are presented in this article utilizing silica and air ($n_{SiO_2} = 1.45$ and $n_{air} = 1$) as the host materials. For both of these designs, silicon ($n_{Si}$=3.45) is considered as the material for the rods. A rod located in the middle of the square unit cell is shown in Fig. 2(a). For $a_{GPC} = 160nm$, the radius of rods in air and silica hosts was calculated based on Eq. (5) to implement the effective refractive index of the unit cell which is illustrated in Fig. 2(b). The GPC-based implementations of the truncated square MFE lens are shown in Fig. 3 where host material is air for Fig. 3(a) and silica for Fig. 3(b). The rods in the corners of the lens were not implemented since they did not affect the performance of the lens.

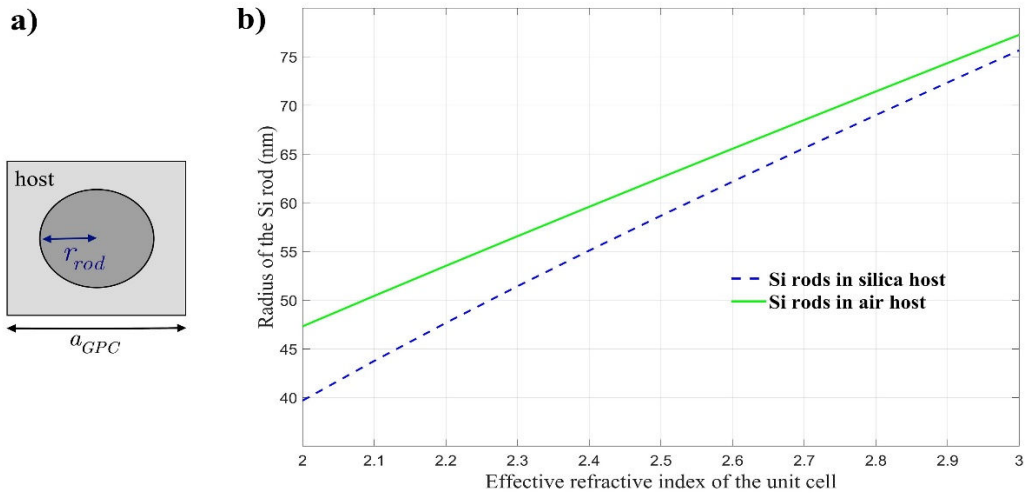

Fig. 2. a) The square unit cell used in the calculations of GPC structure. b) Radius of the silicon rod calculated by Equation 5 to implement the effective refractive index of the unit cell.

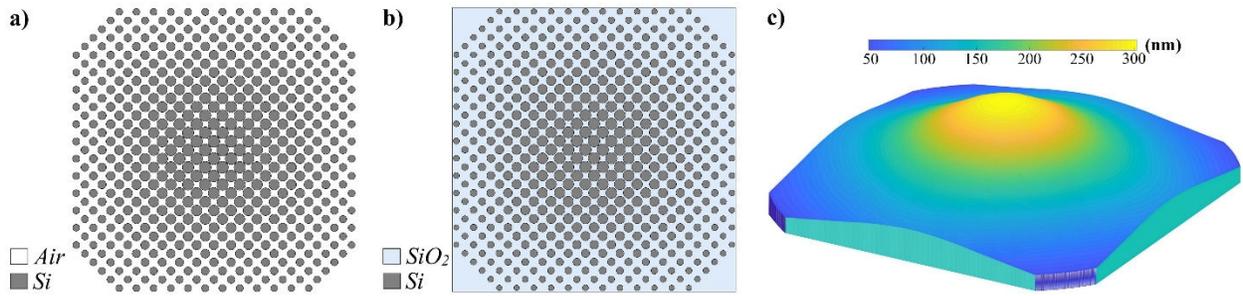

Fig. 3. GPC-based 2D implementation of truncated square MFE with a) air and b) SiO$_2$ as host materials. c) The 3D implementation based on varying silicon thickness. The SiO$_2$ and air claddings are not shown in the 3D implementation.

### 3.2 Varying the thickness of Si slab waveguide

GRIN structures can be implemented by changing the thickness of silicon slab waveguide and consequently changing the effective refractive index of the slab [14, 24, 25]. In order to map the designed refractive index to the height of the guiding layer, a silicon slab waveguide with silica substrate and upper cladding layer of air was considered. The thickness of silica and air layers were assumed to be 2 μm in our calculations. Subsequently, for the given thickness of the Si layer, the effective refractive index of the slab was calculated numerically for the TE mode. Since the calculation method is presented in [14], it is not repeated here. Fig. 3(c) illustrates the lens implemented by this method. The silica substrate and the air cladding are not shown in this figure.

## 4. Results and discussion

The 2D and 3D numerical simulations were carried out to evaluate the performance of the square MFE lens as waveguide crossing. An important factor to be considered is the minimization of reflections from the interface of waveguide and the square MFE lens. To achieve this goal, the refractive index of the square MFE lens at its sides should match with the refractive index of the waveguide's core. To this end, the truncated square lens was chosen in a manner that the boundary of the lens and waveguide had almost equal refractive indices. The width of the waveguides was chosen as $w_{WG} = 1.6 \mu m$ with a thickness of $t_{WG} = 100nm$ to support at least three modes. In order to use the circular MFE lens as crossing medium, the refractive index of the lens's edge should be matched with the waveguides. Consequently, the lens of the virtual domain should be truncated to a circle with a radius of 1.2 μm. The 2×2 waveguide crossing based on the truncated circular MFE lens has been studied in [30]. Our simulations and the results of [30] confirm that the radius of the truncated circular lens should be at least twice the width of the crossing waveguides in order to achieve reasonable results. In our case, the radius of the truncated circular lens is smaller than the waveguide width, therefore, the performance of the virtual domain as the crossing medium is not discussed in this paper. For 2D simulations, the waveguides, with a refractive index equal to the effective refractive index of the slab waveguide (n=2.1), are assumed to be surrounded by air. The refractive index of the designed lens at its interface with the waveguides is not thoroughly constant. The refractive index of the designed lens is 2.1 at the center and is 2.0 at the edges of the waveguides. However, this 0.1 variation of the refractive index at the interface has negligible effect on the performance of the crossing. Our simulations reveal that increasing the size of the lens results in a constant refractive index interface but the improvement in the performance of the crossing is small. For the Fig. 3(b) implementation with SiO$_2$ host material, the electric field distribution of TE$_0$ and TE$_1$ modes are shown in Fig. 4. The simulation results of the TE$_2$ mode are not presented graphically. Due to the imaging properties of the MFE lens, the designed waveguide crossing supports the simultaneous propagation of optical signals with different or same modes from the two waveguides.

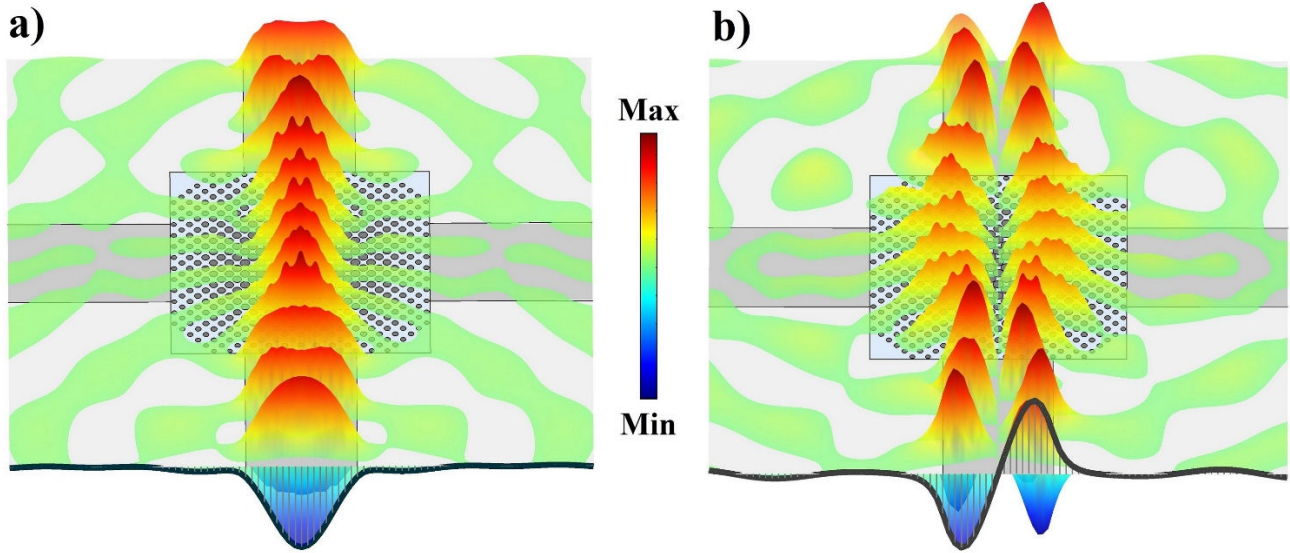

Fig. 4. The electric field distribution of a) TE0 and b) TE1 are shown for GPC-based implementation with SiO2 host material.

The contour plot of the fundamental mode is shown in Fig. 5(a). The transmission, reflection, and crosstalk levels of the fundamental mode for GPC-based (2D) and varying the thickness of Si slab waveguide (3D) implementations of the waveguide crossing are illustrated in Fig. 5(b). The 2D simulation results correspond to the GPC-based implementation with SiO$_2$ host material of Fig. 3(b). In GPC-based implementation, the average insertion loss of 0.7 *dB*, maximum return loss of 23.7 *dB*, and crosstalk levels lower than -47.6 *dB* was achieved for the TE$_0$ mode. And in varying the thickness of Si slab waveguide implementation, the average insertion loss of 0.24*dB*, maximum return loss of 54 *dB*, and crosstalk lower than -72 *dB* was achieved. The contour plot of TE$_1$ mode field profile is shown in Fig. 6(a). The transmission, reflection, and crosstalk levels for 2D (with SiO$_2$ host material) and 3D implementations are shown in Fig. 6(b). For the TE$_1$ mode, the average insertion loss of 0.9 *dB*, maximum return loss of 19.5 *dB*, and crosstalk lower than -26.3 *dB* was achieved for the GPC-based waveguide crossing. The average insertion loss of 0.55 *dB*, maximum return loss of 53 *dB*, and crosstalk lower than -61 *dB* was achieved for the first-order mode, in varying the thickness of Si slab waveguide. The simulation results of the GPC-based implementation with air host material are not presented graphically. For this implementation, the average insertion loss of 0.47 (0.85) *dB* , crosstalk levels lower than -46.1 (-26.7) *dB*, and maximum return loss of 16.1 (13.4) *dB* for TE$_0$ (TE$_1$) mode were achieved. The designed waveguide intersection covers the entire O, E, S, C, L, and U bands of optical communications for both TE$_0$, TE$_1$, and TE$_2$ modes. The designed square MFE lens and the waveguides also support TE$_2$. The electric field distribution of the TE$_2$ mode at 1550nm based on the 3D simulations is illustrated in Fig. 7. The crosstalk levels as well as return and insertion losses are given in this figure for a wavelength of 1550 nm. However, the scattering parameters for this mode are not presented graphically in this paper. For the varying the thickness of the guiding layer implementation, the average insertion loss of 0.45 *dB* , crosstalk lower than -27.3 *dB*, and maximum return loss of 30.1 *dB* were achieved for TE$_2$ mode. Our simulations show that our method can be expanded to support more modes by increasing the lens size and the width of the waveguides. The difference between the result of the GPC-based and varying the thickness of Si slab implementations stems from the fact that the cell's size is finite in the GPC-based implementation while the thickness of silicon layer is gradually varying similar to the ideal lens. Therefore, the performance of the varying the thickness of Si slab implementation is better than the GPC-based implementation. On the other hand, different simulation settings of FEM and FDTD methods, such as perfectly matched layer (PML) settings, may contribute to the differences between the results of these methods.

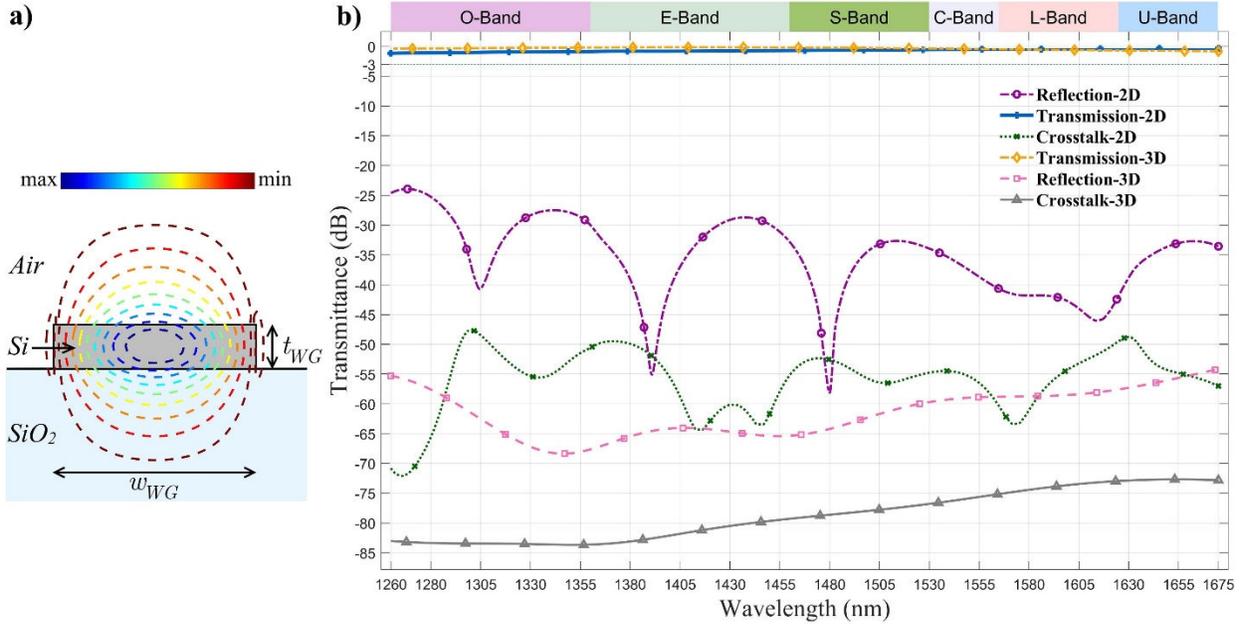

Fig. 5. a) The contour plot of mode field and b) scattering parameters of the 2D GPC-based implementation with SiO$_2$ host material and 3D simulation of varying the thickness of guiding layer implementation are shown for the fundamental mode (TE$_0$).

In order to investigate the similarity of the optical pulse entering the square MFE lens and the pulse leaving the lens we have calculated the fidelity factor (FF) based on the FDTD simulations. The FF is defined as the maximum cross-correlation of the normalized input and output electric field pulses of the crossing:

$$FF = \max_{\tau}\left[\int_{-\infty}^{\infty} \tilde{E}_{in}(t)\tilde{E}_{out}(t+\tau)dt\right] \quad (6)$$

where the delay $\tau$ is varied to maximize the above integral. Moreover, $\tilde{E}_{in}(t)$ and $\tilde{E}_{out}(t)$ are the normalized input and output electric field pulses, respectively:

$$\tilde{E}_{in}(t) = \frac{E_{in}(t)}{\sqrt{\int_{-\infty}^{\infty}|E_{in}(t)|^2 dt}} \quad, \quad \tilde{E}_{out}(t) = \frac{E_{out}(t)}{\sqrt{\int_{-\infty}^{\infty}|E_{out}(t)|^2 dt}} \quad (7)$$

FF is unity when the two signals are identical. The FF were 0.9857, 0.9847, and 0.9817 for TE$_0$, TE$_1$, and TE$_2$ modes. The input optical pulse to the MFE lens and the output pulse are shown in Fig. 8 for the TE$_0$ mode. It takes about 43.4 *fs* for this signal to pass through the waveguide crossing.

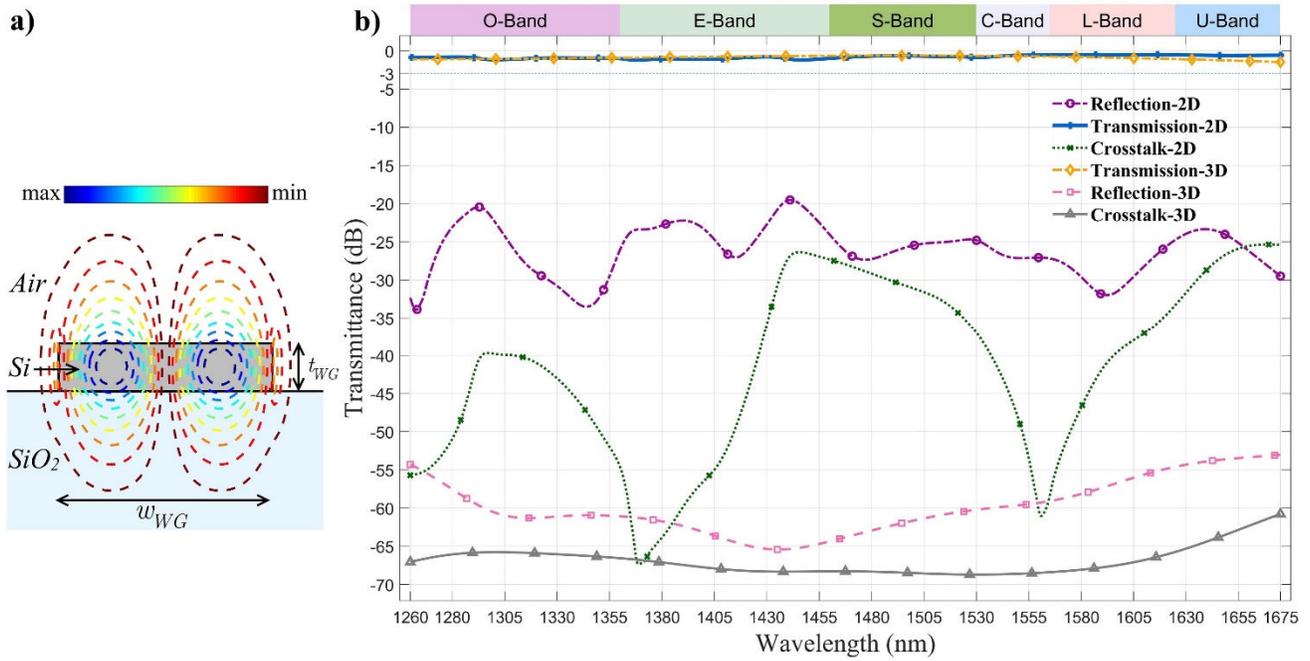

Fig. 6. a) The contour plot of mode field and b) scattering parameters of the 2D GPC-based implementation with $SiO_2$ host material and 3D simulation of varying the thickness of guiding layer implementation are shown for the first-order mode ($TE_1$).

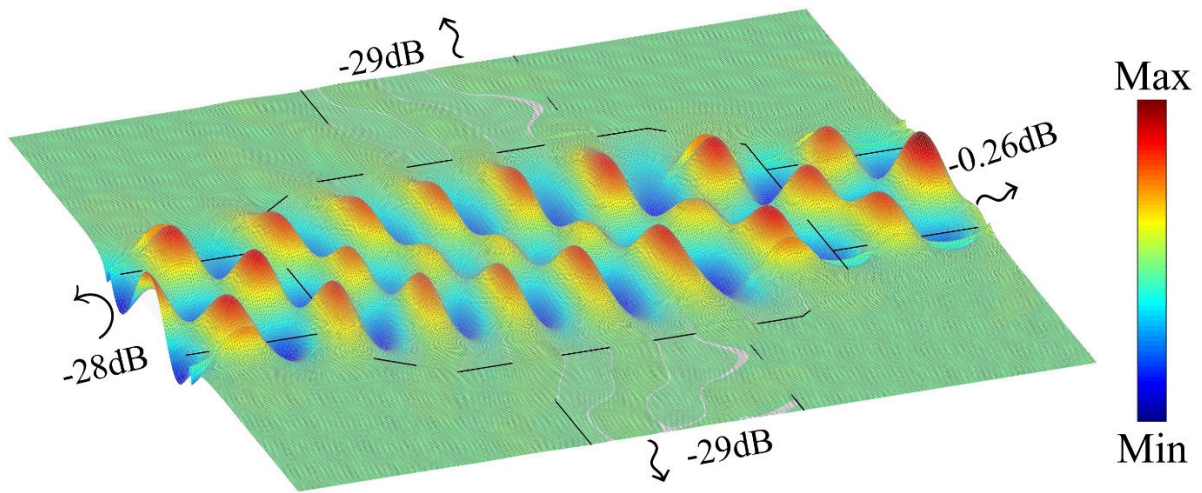

Fig. 7. The $H_z$ component of the $TE_2$ mode at 1550nm. The crosstalk, reflection, and insertion loss are also specified.

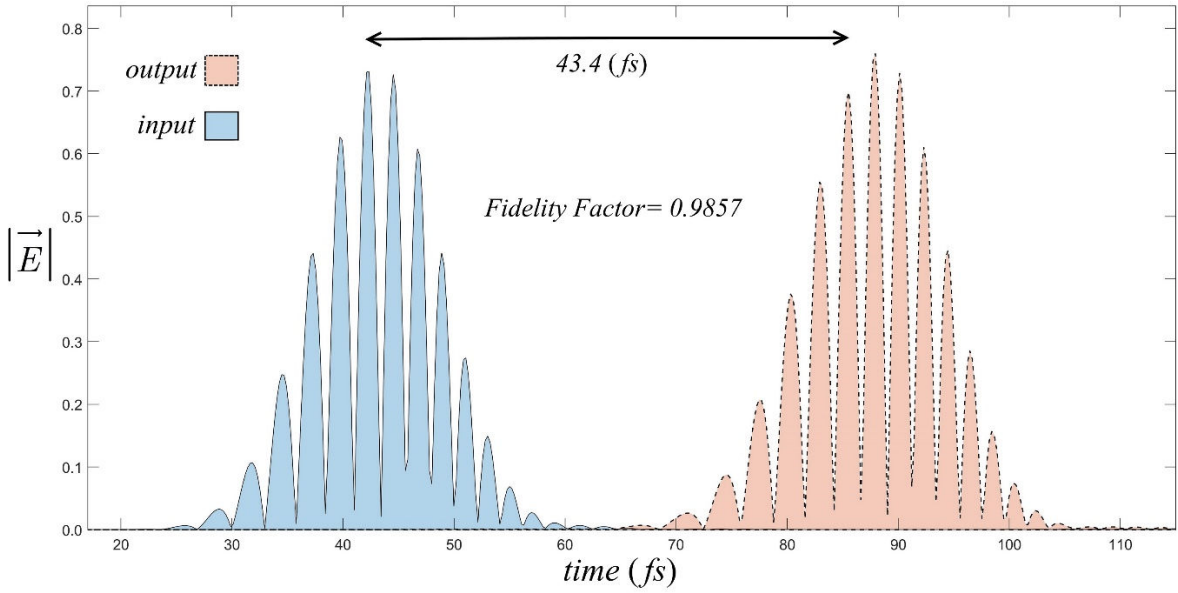

Fig. 8. The optical pulse entering the square MFE lens and the pulse leaving the lens are shown.

### 4.1 Comparison with previous works

The characteristics of the designed multimode intersection and the references of [3-6, 12] are summarized in Table 1. Reference [12] also reports 4×4 waveguide crossing but here we focus on 2×2 waveguide crossings. The crossing mechanism, evaluation method, insertion loss, central wavelength, bandwidth, crosstalk, footprint, and number of supported modes are compared in this table. We should acknowledge that our results are based on numerical simulations while the results reported by [3-6] are based on experimental measurements. For these references, both the simulation and measurement results are presented in the table. To distinguish between the experimental and theoretical results in the table, the theoretical results are in the parentheses. The insertion loss in this table belongs to the highest-order mode supported by the crossing. The overall performance of our design, due to the imaging properties of the MFE lens, is better than other multimode waveguide crossings. Our design has acceptable crosstalk levels and it offers the smallest footprints. This work presents a method to reduce the footprint of the multimode waveguide crossings. Moreover, our design can be easily expanded to support higher number of modes by increasing the size of the lens while crossings of [3, 5, 6] already have large footprints (larger than 21×21 μm$^2$). Compared to our design, reference [4] is the only design with a comparable footprint but it is based on MMI which relies on creating the self-images of the modes at the center of the crossing. Therefore, choosing the width and length of the MMI section becomes complicated as the number of modes increases. Our designed 2×2 waveguide crossing has the broadest bandwidth compared to other studies, covering the entire O, E, S, C, L, and U bands of optical communication. We also introduced random deviations in the thickness of the MFE lens to account for the inevitable fabrication imperfections in its performance. The 10% random thickness deviation introduces an excess insertion loss of up to 0.3 dB.

Table 1. Comparison of the proposed design and other multimode waveguide intersections

| Ref. | Crossing Mechanism | Evaluation method | Insertion Loss (dB) | $\lambda_{center}$ (nm) | Bandwidth (nm) | Cross-talk (dB) | Footprint μm$^2$ | number of supported modes |
|---|---|---|---|---|---|---|---|---|
| [3] | MMI coupler | Experimental (Theoretical) | 1.5 (1.7) | 1560 (1560) | 80 (100) | -18 (-32) | 30×30 | 2 |
| [4] | MMI coupler | Experimental (Theoretical) | 0.6 (0.5) | 1560 (1560) | 60 (80) | -24 (-30) | 4.8×4.8 | 2 |
| [5] | symmetric Y-junction | Experimental (Theoretical) | 1.82 (0.3) | 1555 (1550) | 90 (100) | -18 (-30) | 21×21 | 2 |
| [6] | Asymmetric Y-Junction | Experimental (Theoretical) | 2.0 (1.5) | 1560 (1560) | 60 (80) | -20 (-22) | 34×34 | 3 |
| [12] | MFE lens | (Theoretical) | (0.3) | (1550) | (100) | (-30) | 7.6×7.6 | 2 |
| this work | MFE lens | (Theoretical) | (0.55) | (1467) | (415) | (-27) | 3.77×3.77 | 3 |

## 5. Conclusion

TO provides a mathematical technique for designing a GRIN medium to manipulate the flow of light. We have designed a multimode waveguide crossing based on square MFE lens with QCTO. The designed waveguide crossing has an ultra-wide bandwidth of $415nm$ centered at $1467nm$ with compact footprint of $3.77\mu m \times 3.77\mu m$. The footprint of the truncated square MFE lens is reduced by 54% compared to the circular MFE lens. The 2D FEM simulations were performed to evaluate the performance of the waveguide crossing implemented with GPC. Moreover, the refractive index of a square MFE was mapped to Si thickness distribution and 3D FDTD simulations show that the waveguide crossing had average insertion loss of 0.24, 0.55, and $0.45dB$, and crosstalk lower than -72 , -61, and $-27dB$ for $TE_0$, $TE_1$, and $TE_2$ modes.

## References

1. X. Wu, C. Huang, K. Xu, C. Shu, H.K. Tsang, Mode-division multiplexing for silicon photonic network-on-chip, Journal of Lightwave Technology, 35 (2017) 3223-3228.
2. C. Li, D. Liu, D. Dai, Multimode silicon photonics, Nanophotonics, 8 (2018) 227-247.
3. H. Xu, Y. Shi, Dual-mode waveguide crossing utilizing taper-assisted multimode-interference couplers, Optics letters, 41 (2016) 5381-5384.
4. W. Chang, L. Lu, X. Ren, D. Li, Z. Pan, M. Cheng, D. Liu, M. Zhang, Ultracompact dual-mode waveguide crossing based on subwavelength multimode-interference couplers, Photonics Research, 6 (2018) 660-665.
5. C. Sun, Y. Yu, X. Zhang, Ultra-compact waveguide crossing for a mode-division multiplexing optical network, Optics letters, 42 (2017) 4913-4916.
6. W. Chang, L. Lu, X. Ren, L. Lu, M. Cheng, D. Liu, M. Zhang, An Ultracompact Multimode Waveguide Crossing Based on Subwavelength Asymmetric Y-Junction, IEEE Photonics Journal, 10 (2018) 1-8.
7. D. Headland, W. Withayachumnankul, R. Yamada, M. Fujita, T. Nagatsuma, Terahertz multi-beam antenna using photonic crystal waveguide and Luneburg lens, APL Photonics, 3 (2018) 126105.
8. B. Arigong, J. Ding, H. Ren, R. Zhou, H. Kim, Y. Lin, H. Zhang, Design of wide-angle broadband luneburg lens based optical couplers for plasmonic slot nano-waveguides, Journal of Applied Physics, 114 (2013) 144301.
9. M.M. Gilarlue, S.H. Badri, H. Rasooli Saghai, J. Nourinia, C. Ghobadi, Photonic crystal waveguide intersection design based on Maxwell's fish-eye lens, Photonics and Nanostructures - Fundamentals and Applications, 31 (2018) 154-159.
10. S.H. Badri, M.M. Gilarlue, Maxwell's fisheye lens as efficient power coupler between dissimilar photonic crystal waveguides, Optik, 185 (2019) 566-570.
11. S.H. Badri, M.M. Gilarlue, Low-index-contrast waveguide bend based on truncated Eaton lens implemented by graded photonic crystals, J. Opt. Soc. Am. B, 36 (2019) 1288-1293.
12. H. Xu, Y. Shi, Metamaterial-Based Maxwell's Fisheye Lens for Multimode Waveguide Crossing, Laser & Photonics Reviews, 1800094.
13. M.M. Gilarlue, J. Nourinia, C. Ghobadi, S.H. Badri, H.R. Saghai, Multilayered Maxwell's fisheye lens as waveguide crossing, Optics Communications, 435 (2019) 385-393.
14. S.H. Badri, H.R. Saghai, H. Soofi, Polygonal Maxwell's fisheye lens via transformation optics as multimode waveguide crossing, Journal of Optics, (2019). https://doi.org/10.1088/2040-8986/ab1dba
15. O. Quevedo-Teruel, W. Tang, R.C. Mitchell-Thomas, A. Dyke, H. Dyke, L. Zhang, S. Haq, Y. Hao, Transformation optics for antennas: why limit the bandwidth with metamaterials?, Scientific reports, 3 (2013) 1903.
16. D.H. Werner, D.-H. Kwon, Transformation electromagnetics and metamaterials, Springer2014.
17. [17] R. Duraiswami, A. Prosperetti, Orthogonal mapping in two dimensions, Journal of Computational Physics, 98 (1992) 254-268.
18. J. Li, J.B. Pendry, Hiding under the carpet: a new strategy for cloaking, Physical review letters, 101 (2008) 203901.
19. Q. Wu, J.P. Turpin, D.H. Werner, Integrated photonic systems based on transformation optics enabled gradient index devices, Light: Science & Applications, 1 (2012) e38.
20. L. Xu, H. Chen, Conformal transformation optics, Nature Photonics, 9 (2015) 15.
21. E. Centeno, D. Cassagne, Graded photonic crystals, Optics letters, 30 (2005) 2278-2280.
22. U. Leonhardt, T.G. Philbin, Transformation optics and the geometry of light, Progress in Optics, 53 (2009) 69-152.
23. B. Vasić, G. Isić, R. Gajić, K. Hingerl, Controlling electromagnetic fields with graded photonic crystals in metamaterial regime, Optics Express, 18 (2010) 20321-20333.
24. A. Di Falco, S.C. Kehr, U. Leonhardt, Luneburg lens in silicon photonics, Optics express, 19 (2011) 5156-5162.
25. O. Bitton, R. Bruch, U. Leonhardt, Two-dimensional Maxwell fisheye for integrated optics, Physical Review Applied, 10 (2018) 044059.